\documentclass[aps,twocolumn,showpacs,superscriptaddress]{revtex4-1}

\usepackage[T1]{fontenc}
\usepackage{amsfonts,amsmath,amsbsy,amssymb}
\usepackage{graphicx,color}
\usepackage{times}

\let\mathbf=\boldsymbol

\def\blue#1{\textcolor{black}{#1}}

\begin{document}

\title{High-topological-number magnetic skyrmions and topologically protected dissipative structure}

\author{Xichao Zhang}
\affiliation{Department of Physics, University of Hong Kong, Hong Kong, China}
\affiliation{School of Electronics Science and Engineering, Nanjing University, Nanjing 210093, China}
\author{Yan Zhou}
\email[]{yanzhou@hku.hk}
\affiliation{Department of Physics, University of Hong Kong, Hong Kong, China}
\affiliation{School of Electronics Science and Engineering, Nanjing University, Nanjing 210093, China}
\author{Motohiko Ezawa}
\email[]{ezawa@ap.t.u-tokyo.ac.jp}
\affiliation{Department of Applied Physics, University of Tokyo, Hongo 7-3-1, 113-8656, Japan}

\begin{abstract}
The magnetic skyrmion with the topological number of unity ($Q=1$) is a well-known nanometric swirling spin structure in the nonlinear $\sigma$ model with the Dzyaloshinskii-Moriya interaction.
Here, we show that magnetic skyrmion with the topological number of two ($Q=2$) can be created and stabilized by applying vertical spin-polarized current though it cannot exist as a static stable excitation.
Magnetic skyrmion with $Q=2$ is a nonequilibrium dynamic object, subsisting on a balance between the energy injection from the current and the energy dissipation by the Gilbert damping.
Once it is created, it becomes a topologically protected object against fluctuations of various variables including the injected current itself.
Hence, we may call it a topologically protected dissipative structure.
We also elucidate the nucleation and destruction mechanisms of the magnetic skyrmion with $Q=2$ by studying the evolutions of the magnetization distribution, the topological charge density as well as the energy density.
Our results will be useful for the study of the nontrivial topology of magnetic skyrmions with higher topological numbers.
\end{abstract}


\date{\today}

\maketitle

\section{Introduction}
There is a long history of skyrmions from the particle physics to the condensed matter physics~\cite{Skyrme61,Belavin75,Rajaraman,MultiSkyrmion}.
Originally, Skyrme introduced "skyrmions" in the three-dimensional (3D) space to describe nucleons as elementary particles possessing nontrivial topological numbers~\cite{Skyrme61}.
Subsequently, Belavin and Polyakov (BP) applied the concept to the two-dimensional (2D) ferromagnetic (FM) system and predicted a magnetic excitation carrying a nontrivial topological number, that is, the Pontryagin number~\cite{Belavin75}.
The BP-skyrmion is an exact solution of the nonlinear $\sigma$ model. However this solution has a scale invariance, and hence the BP-skyrmion has no definite radius. It is necessary for a dynamically stable and physical skyrmion to have a definite radius.
Consequently, a skyrmion is characterized by two types of stabilities: the topological stability based on the conservation of a nontrivial Pontryagin number, and the dynamical stability having a finite radius.

Some interactions are required in order to break the scale invariance. For instance, the dipole-dipole interaction breaks the scale invariance. However, since it is a long-distance force, the skyrmion size becomes very large in general~\cite{Ezawa,Marco}.
Recently, the Dzyaloshinskii-Moriya interaction (DMI) has attracted much attention to provide the skyrmion with a finite radius, where the resultant radius is of the order of nanometers~\cite{Bogdanov,SkRev,Fert,Sampaio,Mol,Yu,Heinze,Schulz,YuRoom,Munzer}.
It is called the magnetic skyrmion.
Magnetic skyrmions might be suitable for building next-generation nonvolatile memory devices based on their topological stability~\cite{Sampaio}. Furthermore, it has also been proposed to utilize them in logic computing~\cite{XichaoSR2015}.

Strictly speaking, the topological number is defined only in the continuum field theory with an infinitely large space.
Although the underlying system has a finite size and a lattice structure in the condensed matter physics, the topological number is well-defined, provided that the skyrmion spin texture is sufficiently smooth and sufficiently far away from the edge.
Thus it is an intriguing concept in the condensed matter physics possessing the aspect of the continuum theory and that of the lattice theory. We can actually leverage these properties for practical applications. It is possible to create or destroy magnetic skyrmions which are topologically stable~\cite{Ezawa,Marco,Tchoe,Iwasaki,NC,Yan,Romming,Lin}.

As far as the topological analysis is concerned, a magnetic skyrmion with any topological number $Q$ is possible. Let us call such a skyrmion with $Q\geq 2$ a high-Q skyrmion.
It will be realized when the in-plane component of the spin rotates by $2\pi Q$ and the skyrmion acquires a high helicity $Q$.
However, only the magnetic skyrmion with $Q=1$ has so far been realized. This is because that a static magnetic skyrmion with $Q\geq 2$ is unstable since the DMI cannot prevent it from shrinking to a point, which we shall prove later.
In this paper, employing a dynamical breaking of the scale invariance~\cite{Yan}, we create a high-Q skyrmion with $Q=2$ by applying a spin-polarized current perpendicular to the FM nanodisk with the DMI. We investigate in details how a high-Q skyrmion with $Q=2$ is created from a magnetic bubble with $Q=0$ through successive creations of two Bloch points.

\begin{figure*}[t]
\centerline{\includegraphics[width=1.00\textwidth]{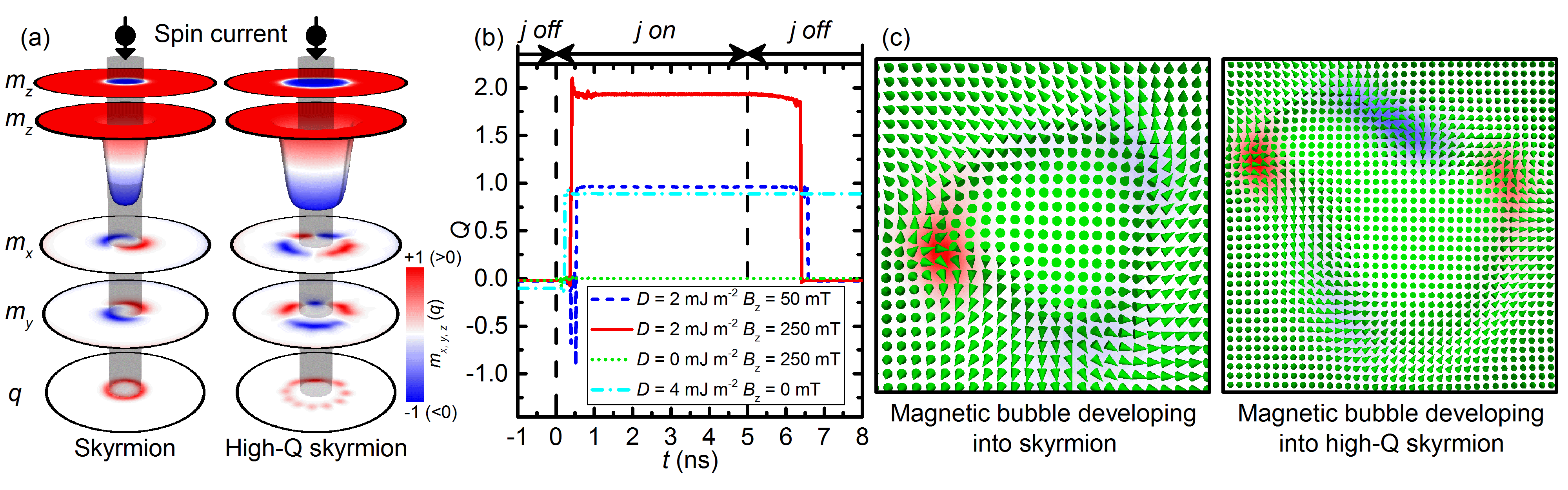}}
\caption{
(Color online)
The ordinary skyrmion ($Q=1$) and the high-Q skyrmion ($Q=2$).
(a) Our system consists of a $1$-nm-thick FM nanodisk with a thickness of $150$ nm, where the spin current polarized along the $-z$-direction is injected into the central $30$-nm-diameter region. The out-of-plane magnetization $m_z$, the in-plane magnetization ($m_x, m_y$) and the topological charge density $q$ are illustrated for the skyrmion ($Q=1$) and the high-Q skyrmion ($Q=2$).
(b) The topological number $Q$ as a function of time $t$. We create a skyrmion with $Q=1$ ($Q=2$) by applying an external magnetic field perpendicular to the nanodisk plane pointing along the $+z$-direction with an amplitude of $B_{z}=50$ mT ($B_{z}=250$ mT), as depicted in blue (red).
The DMI constant is $D=2$ mJ m$^{-2}$. The spin current with a current density of $j=3\times 10^{12}$ A m$^{-2}$ is switched on and off at $t=0$ and $t=5$ ns.
When the DMI is set to be $D=0$ mJ m$^{-2}$, no skyrmion is formed ($Q=0$), as depicted in green.
A skyrmion with $Q=1$ remains stable as it is even after the spin current is switched off when $D=4$ mJ m$^{-2}$ and $B_{z}=0$ mT, as depicted in cyan, which is consistent with the previous result shown in Ref.~\onlinecite{Sampaio}.
(c) A magnetic bubble with $Q=0$ has a pair (two pairs) of blue and red areas, when it develops into a skyrmion with $Q=1$ ($Q=2$). The cones represent the magnetization, while the color denotes the topological charge density.
}
\label{FIG1}
\end{figure*}

Once the spin texture of a high-Q skyrmion becomes sufficiently smooth with respect to the lattice spacing, the topological protection becomes active. Namely, the resultant spin texture is topologically robust against fluctuations of various variables including the injected current itself. This is due to the fact that the topological number cannot change continuously from its quantized value. However, when the current is switched off, the high-Q skyrmion quickly shrinks to the order of the lattice scale and dissipates. The equilibrium and dissipation processes are well-described in terms of the Rayleigh dissipation function composed of the energy injection and dissipation terms~\cite{Goldstein}. The dissipation is found to spread all over like a burst at the moment of the skyrmion generation and destruction~\cite{burst}.

The high-Q skyrmion is a topologically protected dissipative structure. One might think that the concept is a self-contradictory proposition. However, this is not the case.
We may suggest an analogy of the quantum Hall (QH) state. It is well-known that the QH state at an integer filling factor is robust against impurities and also against the change of the external magnetic field. Indeed the latter develops QH plateaux. The robustness is due to the fact that the state is protected by the conservation of the topological number, that is, the Chern number. Nonetheless, when the external magnetic field is switched off, the topological robustness is lost and the QH state collapses. Clearly, the external magnetic field in the QH system corresponds to the injected spin current in the topologically stabilized dissipative structure of this work.

\section{Numerical results on the high-Q skyrmion}
\subsection{Nucleation of the high-Q skyrmion}
Our system is composed of a FM nanodisk and a spin-polarized current injection region with a radius of $r_{c}$, as illustrated in Fig.~\ref{FIG1}(a) (see the Appendix for modeling details). The development of the topological number $Q$ is shown in Fig.~\ref{FIG1}(b). Soon after the spin current is injected, $Q$ suddenly increases to $1$ or $2$ from $0$, and remains stable, in the presence of the DMI. On the other hand, when the spin current is switched off, $Q$ decreases to $0$.
It should be noted that the high-Q skyrmion cannot be created dynamically without the DMI.

\begin{figure*}[t]
\centerline{\includegraphics[width=1.00\textwidth]{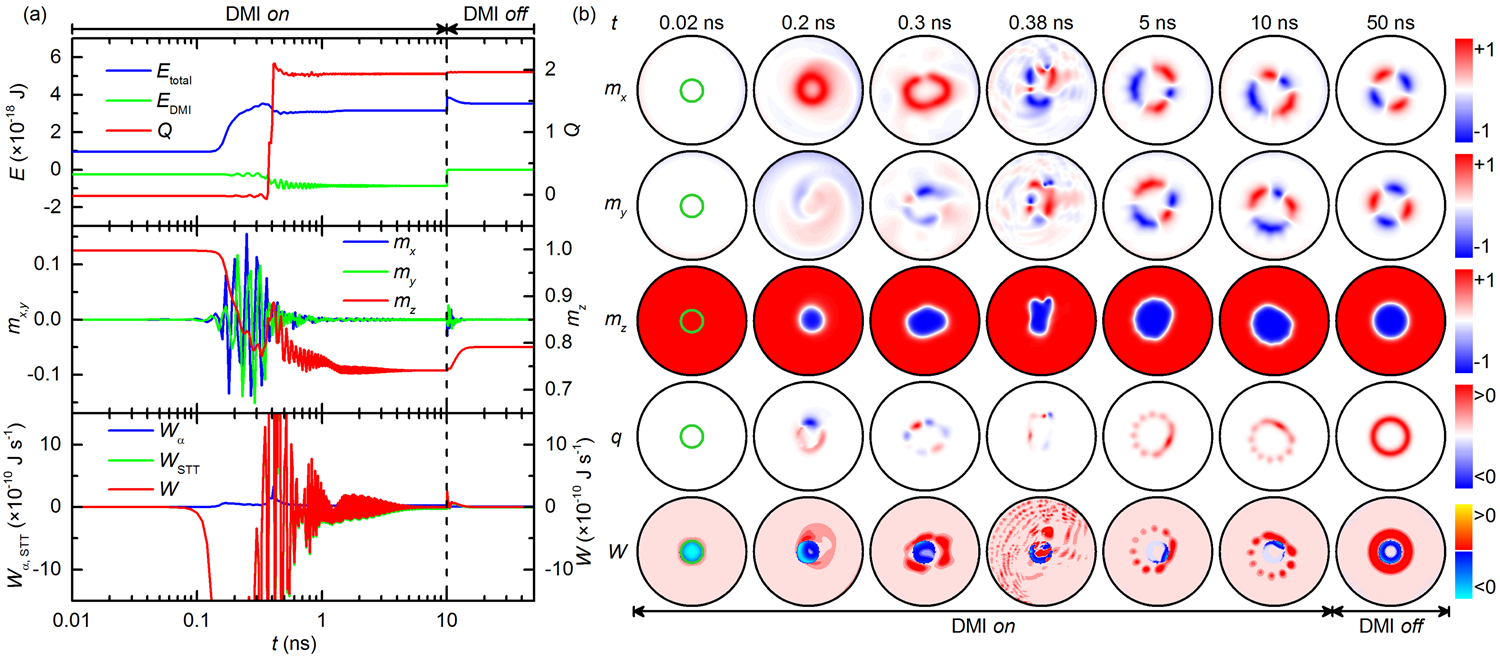}}
\caption{
(Color online)
Time evolution of the high-Q skyrmion ($Q=2$).
(a) Time evolution of the total energy $E_{\text{total}}$, the DMI energy $E_{\text{DMI}}$, the topological number $Q$, the in-plane ($m_x, m_y$) and out-of-plane $m_{z}$ components of magnetization averaged over the simulation area, and the dissipation functions $W$. A high-Q skyrmion with $Q=2$ is created nearly at $t=0.38$ ns.
The DMI constant $D=2$ mJ m$^{-2}$. The spin current density $j=3\times 10^{12}$ A m$^{-2}$. The external magnetic field $B_z=250$ mT.
(b) Top-views of the magnetization distributions $m_{x}$, $m_{y}$ and $m_{z}$ of the FM nanodisk, the corresponding topological charge density distribution $q$ and the Rayleigh dissipation function $W$ at selected times. The green circle indicates the spin current injection region. The nucleation of a high-Q skyrmion with $Q=2$ occurs nearly at $t=0.38$ ns, where the dissipation function spreads all over the FM nanodisk, implying that the spin wave propagates. The DMI is turned off at $t=10$ ns. It is remarkable that the high-Q skyrmion remains stable even if the DMI is switched off, which demonstrates the topological protection against the change of a variable, that is, the DMI.
}
\label{FIG2}
\end{figure*}

We are interested in the process how a magnetic bubble with $Q=0$ is converted into a magnetic skyrmion with $Q\not=0$. Upon the application of the spin current, there is a large energy injected into the core through the spin transfer torque (STT). The spins are forced to reverse within the core upon the spin-polarized current injection. Due to the DMI, the spins are twisted around the core. The topological number is zero for such a state. This is a magnetic bubble with $Q=0$. 

We point out that the seed of a magnetic skyrmion is already present in the magnetic bubble. In Fig.~\ref{FIG1}(c), we show the densities of the in-plane components of the magnetization ($m_{x}(\mathbf{x}), m_{y}(\mathbf{x})$) and the topological charge density $q(\mathbf{x})$ of a magnetic bubble before the nucleation to a magnetic skyrmion with $Q=1$ or $2$. We clearly observe a pair (two pairs) of blue and red areas indicating negative and positive topological charge densities, respectively. 

In Fig.~\ref{FIG2}(a), we show the time evolution of the topological number, the total energy, the DMI energy, the average magnetization ($m_{x}$, $m_{y}$, $m_{z}$), and the Rayleigh dissipation function $W$ for a high-Q skyrmion with $Q=2$ (\blue{see Ref.~\onlinecite{SI} for Supplementary Movie 1}). The selected top-views are shown in Fig.~\ref{FIG2}(b). The spin component $m_{z}$ measures the size of the skyrmion, while $m_{x}$ and $m_{y}$ contribute to the topological charge density. First, $m_{z}$ starts to decrease, implying that the spins are inverted in the disk region. However, the topological number remains zero, since the in-plane spin components point along the same direction, as shown in Fig.~\ref{FIG2}(b) at $t=0.2$ ns.

The Rayleigh dissipation function gives us a vivid information on how the dissipation occurs in the dissipative system. In the present system, the energy is injected into the core and dissipated in its outer side steadily before and after the nucleation of a magnetic skyrmion. However, the dissipation spreads around all over the FM nanodisk like a burst at the transition moment of the topological number from $Q=0$ to $Q=2$ (see Fig.~\ref{FIG2}(b) at $t=0.38$ ns).

\begin{figure}[t]
\centerline{\includegraphics[width=0.50\textwidth]{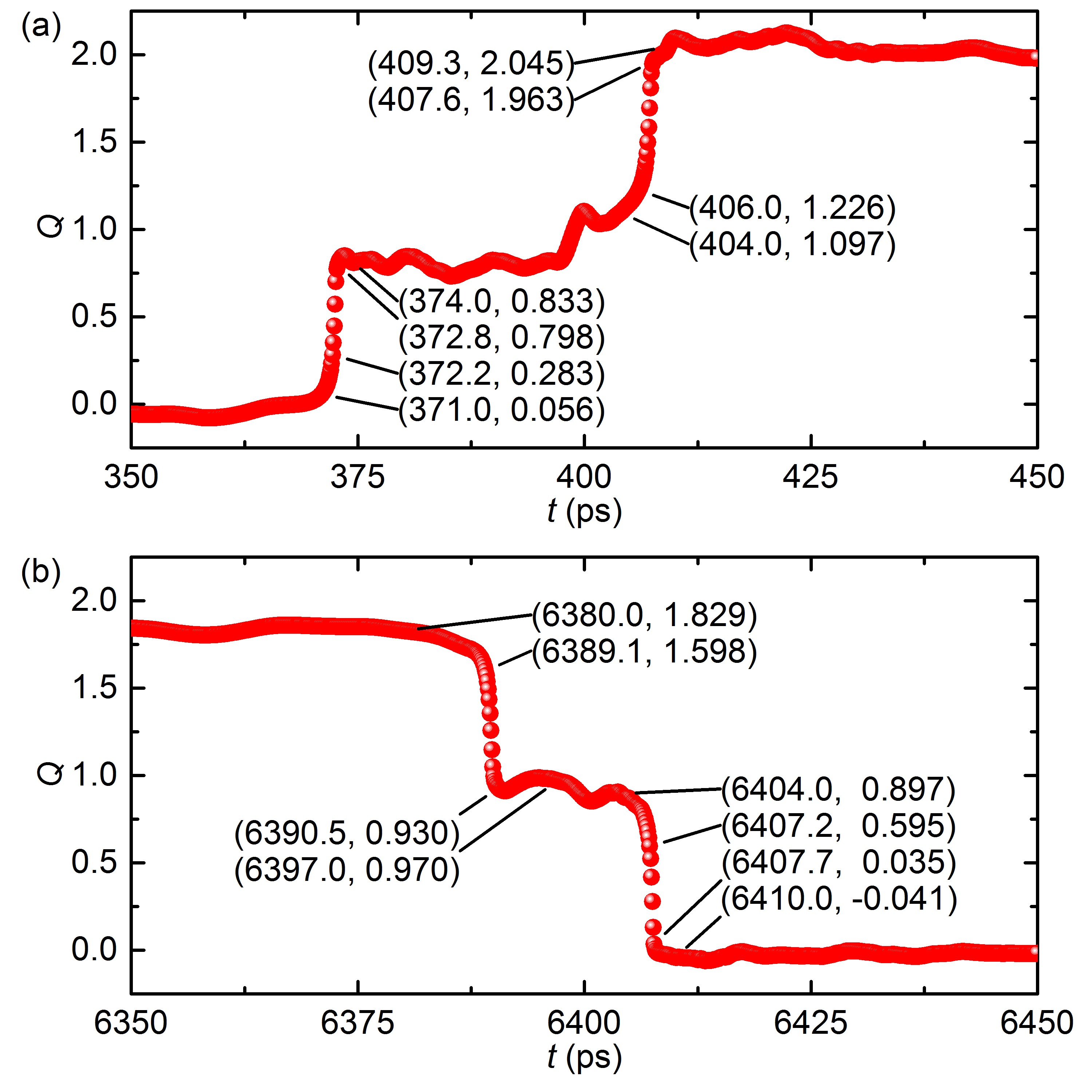}}
\caption{
(Color online)
(a) Nucleation and (b) annihilation processes of the high-Q skyrmion with $Q=2$. A sudden change of the topological number $Q$ occurs twice successively when the high-Q skyrmion with $Q=2$ is created or destroyed. The topological charge density $q(\mathbf{x})$, the energy density $\epsilon(\mathbf{x})$, and the spin-component distribution $m_z(\mathbf{x})$ of the state indexed by $(t, Q)$ in (a) and (b) are shown in Fig.~\ref{FIG4} and Fig.~\ref{FIG9}, respectively.
}
\label{FIG3}
\end{figure}

\begin{figure*}[t]
\centerline{\includegraphics[width=1.00\textwidth]{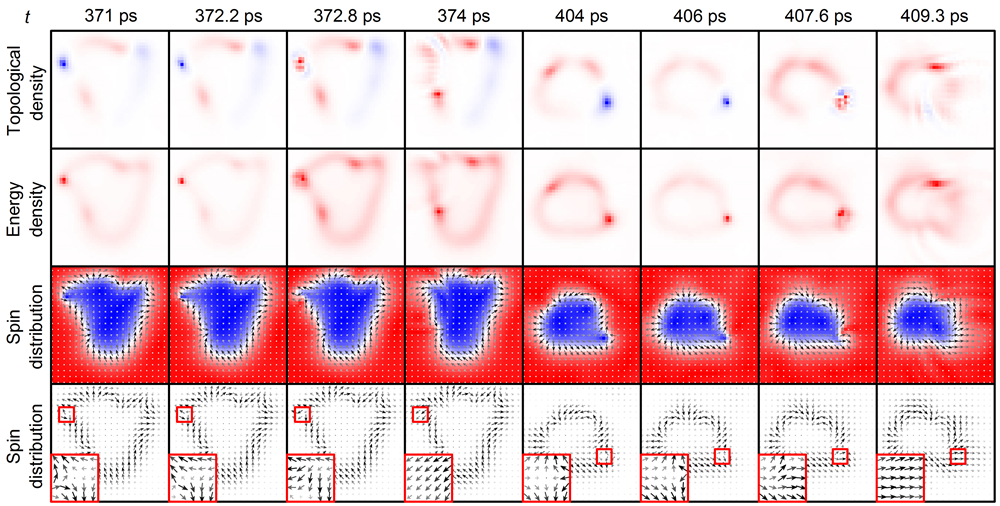}}
\caption{
(Color online)
Nucleation process of the high-Q skyrmion with $Q=2$.
Snapshots of the topological charge density $q(\mathbf{x})$, the energy density $\epsilon(\mathbf{x})$, the spin-component distribution $m_z(\mathbf{x})$, and its close up at sequential times.
The DMI constant $D=2$ mJ m$^{-2}$. The spin current density $j=3\times 10^{12}$ A m$^{-2}$. The external magnetic field $B_z=250$ mT.
In the simulation, each cell corresponds to one spin, and the cell size is $1.5$ nm $\times$ $1.5$ nm $\times$ $1$ nm.
In the spin distribution panels, each arrow stands for four spins, while it stands for one spin in the insets.
The nucleation process of the high-Q skyrmion with $Q=2$ is found to occur in two steps. First, it starts when a high-energy-density part is localized to a lattice-scale area, which possesses almost $Q=-1$.
A few spins rotate by large angles in this area, making the topological number of the area almost zero. The resultant spin texture has $Q=1$.
Second, a similar phenomenon occurs, yielding the high-Q skyrmion with $Q=2$ after the relaxation.
\blue{See Ref.~\onlinecite{SI} for Supplementary Movie 4.}
}
\label{FIG4}
\end{figure*}

We show a close-up of the change of the topological number $Q$ around $t=0.38$ ns in Fig.~\ref{FIG3}(a). Clearly, there are two successive jumps of $Q$ as $0\rightarrow1\rightarrow2$. We also show the topological charge density, the energy density, and the spin distribution around $t=0.38$ ns in Fig.~\ref{FIG4}.

Based on these we obtain the following picture of the nucleation process. We focus on the case where there are two pairs of blue and red areas in the domain boundary region of the magnetic bubble (Fig.~\ref{FIG1}(c)). Both the topological charge density $q(\mathbf{x})$ and the energy density $\epsilon(\mathbf{x})$ are large in these areas. In particular, there is a chance that $\epsilon(\mathbf{x})$ becomes large in a lattice-scale area so that the area has almost $Q=-1$. This happens when two spins are antiparallel with one down-spin site between them, at $t=372.2$ ps in Fig.~\ref{FIG4}. A single or a few spins make a large rotation in order to decrease the energy of the area. Indeed, the antiparallel spins become parallel at $t=372.8$ ps in Fig.~\ref{FIG4}. In this process the topological number $Q=-1$ is lost, which is possible in the lattice theory. This phenomenon would be viewed as a generation of a Bloch point in the continuum theory. This makes clearer the role of the Bloch point presented by Sampaio \textit{et al}. in Ref.~\onlinecite{Sampaio}. There exists still a pair of blue and red areas. Now, there is a chance that another lattice-scale area develops which has $Q=-1$ in the spin texture with $Q=1$. (Note that there are two spins antiparallel at $t=406$ ps, which become parallel at $t=407.6$ ps.) It corresponds to the emergence of another Bloch point. As a result, it turns out that two Bloch points emerge successively in a single magnetic bubble. When the spin texture becomes sufficiently smooth, it yields a high-Q skyrmion with $Q=2$.

\begin{figure}[t]
\centerline{\includegraphics[width=0.50\textwidth]{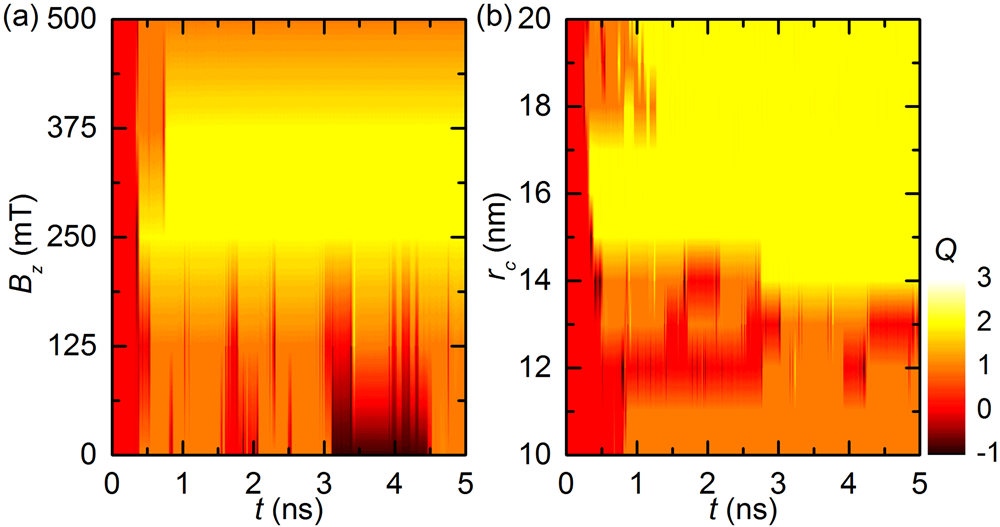}}
\caption{
(Color online)
(a) Phase diagram of the high-Q skyrmion creation with respect to the external magnetic field $B_{z}$ and time $t$.
The spin current density $j=3\times 10^{12}$ A m$^{-2}$, which is injected into a circle region with a radius of $r_{c}=15$ nm. The DMI constant $D=2$ mJ m$^{-2}$.
(b) Phase diagram of the high-Q skyrmion creation with respect to the spin current injection region radius $r_{c}$ and time $t$.
The spin current density $j=3\times 10^{12}$ A m$^{-2}$, which is injected into a circle region with a radius of $r_{c}$. The DMI constant $D=2$ mJ m$^{-2}$. The external magnetic field $B_z=250$ mT.
The color scale indicates the topological number $Q$.
}
\label{FIG5}
\end{figure}

We have also numerically observed that a high-Q skyrmion can be successfully created in a wide range of  parameters as well as the spin current injection size, as shown for instance in Fig.~\ref{FIG5}.
Fig.~\ref{FIG5}(a) shows the phase diagram of the high-Q skyrmion creation with respect to the external magnetic field $B_{z}$ and time, where the spin current is injected into a circle region with a radius of $r_{c}=15$ nm.
Fig.~\ref{FIG5}(b) shows the phase diagram of the high-Q skyrmion creation with respect to the spin current injection region radius $r_{c}$ and the time $t$, where the spin current is injected into a circle region with a radius of $r_{c}$.

\begin{figure}[t]
\centerline{\includegraphics[width=0.50\textwidth]{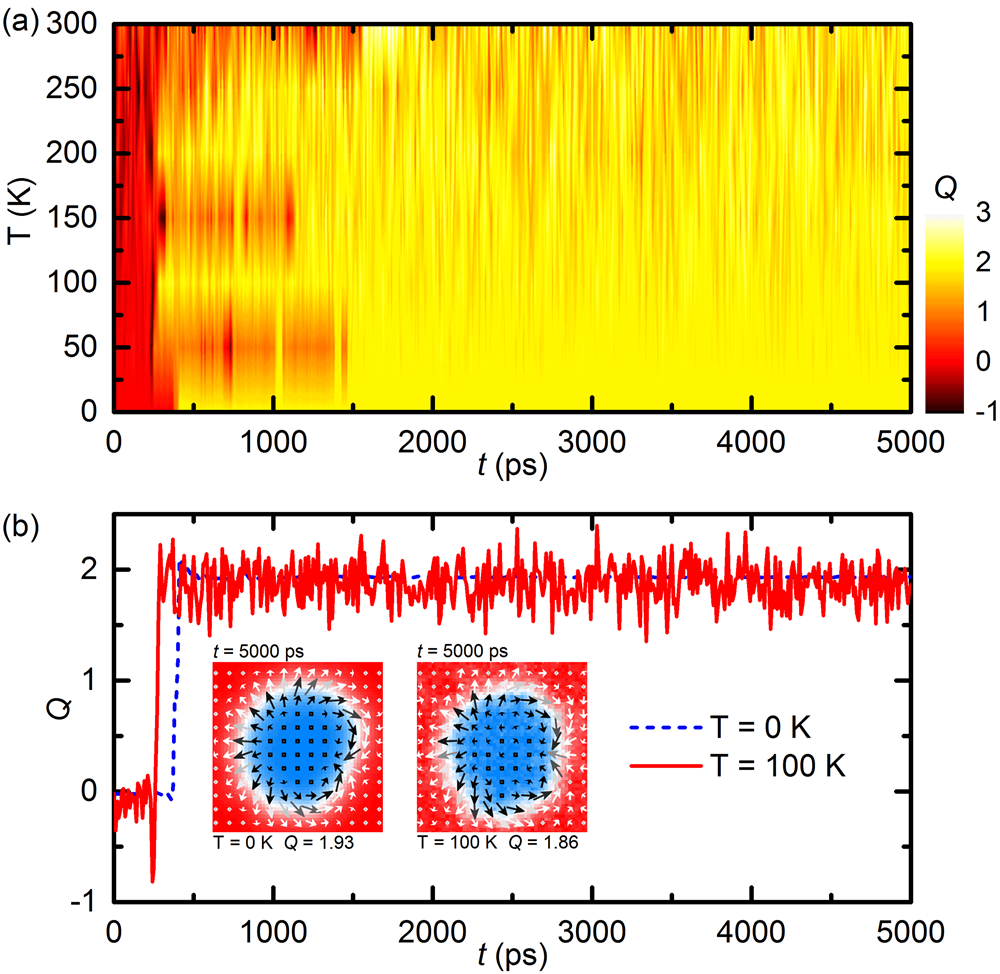}}
\caption{
(Color online)
(a) Phase diagram of the high-Q skyrmion creation with respect to the temperature $\text{T}$ and the time $t$.
The DMI constant $D=2$ mJ m$^{-2}$. The spin current density $j=3\times 10^{12}$ A m$^{-2}$. The external magnetic field $B_z=250$ mT.
The color scale indicates the topological number $Q$.
(b) The topological number $Q$ as a function of time $t$ at $\text{T}=0$ K and $\text{T}=100$ K. The insets show the snapshots of the high-Q skyrmion at $\text{T}=0$ K and $\text{T}=100$ K.
}
\label{FIG6}
\end{figure}

Moreover, as shown in Fig.~\ref{FIG6}, the high-Q skyrmion is created and maintained even at finite temperature.
Fig.~\ref{FIG6}(a) illustrates the phase diagram of the high-Q skyrmion creation with respect to the temperature $\text{T}$ and time.
Fig.~\ref{FIG6}(b) shows the topological number $Q$ as a function of the time at $\text{T}=0$ K and $\text{T}=100$ K, where the high-skyrmion with $Q=2$ is created shortly after the spin current is switched on.
The structure of the high-Q skyrmion is deformed at finite temperature (see Fig.~\ref{FIG6}(b) insets).
Its topological number is almost $2$ but fluctuates slightly because the continuity of the spin texture of a deformed skyrmion is broken at finite temperature.

\subsection{Evolution of the high-Q skyrmion}
The continuity of the spin texture is recovered since a smooth texture has a lower energy. Then the topological protection becomes active. The dissipation function decreases rapidly and oscillates around zero. The system is relaxed to a steady state around $t=5$ ns. Both the topological charge density and the Rayleigh dissipation function are almost zero outside the domain wall encircling the high-Q skyrmion (see Fig.~\ref{FIG2}). 

Oscillations in various variables occur due to the DMI. It is instructive to switch off the DMI in numerical simulations. In Fig.~\ref{FIG2}, we switch off the DMI at $t=10$ ns. First of all, the high-Q skyrmion remains stable. Second, the oscillations in the energy, the magnetization components, and the dissipation functions disappear. After the relaxation at $t=20$ ns, the in-plane magnetization components and the dissipation functions becomes exactly zero (\blue{see Ref.~\onlinecite{SI} for Supplementary Movie 2}). The spin texture is described precisely by the magnetization components of the domain wall.

We show how the balance holds between the energy injection from the spin current and the dissipation by the Gilbert damping in Fig.~\ref{FIG7}. The energy injection occurs in the vicinity of the edge of the current injection region ($r\lesssim r_{c}$) while the dissipation occurs mostly within the domain wall encircling it ($r\sim r_{0}$) together with energy flow from the inner to outer regions. This is due to the fact that spins precess within the domain wall.

\subsection{Topological protection of the high-Q skyrmion}
It is important to point out that this energy balance takes place automatically so as to keep the topological number unchanged. To check this, we change the spin current stepwise, which is shown in Fig.~\ref{FIG8}. The topological number remains to be $2$ when the spin current intensity changes even more than twice. The energy injection due to the STT becomes larger as the spin current density increases. Accordingly, the energy dissipation due to the Gilbert damping increases. As a result, the Rayleigh dissipation function remains zero in average although it is oscillating. The total energy increases stepwise but remains almost constant for each current strength. The total $m_z$ decreases as the spin current density increases, which implies that the magnetic skyrmion expands for larger current density. We have shown that the magnetic skyrmion is topologically robust against a considerable change of the spin current injection.

As we have stated, once it is created, the high-Q skyrmion remains stable even if the DMI is switched off (see Fig.~\ref{FIG2}). Furthermore, it is stable against the fluctuations of various variables. This is because a small change can induce only a small change of the topological number $Q$, but this is impossible since $Q$ is a quantized quantity. This property is called the topological protection. 

\subsection{Destruction of the high-Q skyrmion}
When the spin current is switched off at $t=5$ ns, the topological number remains as $Q=2$ until $t\sim 6.4$ ns and suddenly decreases to $Q=0$, as shown in Fig.~\ref{FIG1}(b). The radius of the magnetic skyrmion shrinks since the skyrmion core is fixed by the spin current against the shrinking force due to the kinetic energy as well as the external magnetic field (\blue{see Ref.~\onlinecite{SI} for Supplementary Movie 3}).

A close examination shows that the collapse of the topological number occurs in two steps as $Q=2\rightarrow 1\rightarrow 0$ as in Fig.~\ref{FIG3}(b). We may understand how the destruction of a magnetic skyrmion with $Q=2$ occurs by investigating the time evolution of the topological charge density, the energy density, and the spin distribution around $t=6.4$ ns as shown in Fig.~\ref{FIG9}.

In the first step the energy density is localized almost on one lattice site, where a Bloch point is generated and the topological number changes from $Q=2$ to $Q=1$. The size of the magnetic skyrmion with $Q=1$ shrinks almost to the order of the lattice site.

The second step has some new features. The topological stability of the magnetic skyrmion is guaranteed by the fact that the core spin points in the direction opposite to the FM background. However, such a core spin disappears at $t=6407$ ps since there is no lattice site at the core spin. As a result, the spin texture becomes a vortex structure. Accordingly, all the spins point along upward direction and the magnetic skyrmion disappears. This is possible since the system is on the lattice and such a transition never happens for the continuum system. In this process, the spiral spin wave is generated, as seen obviously in the topological density as well as the energy density at $t=6410$ ps in Fig.~\ref{FIG9}.

A comment is in order with respect to the stability of a magnetic skyrmion with $Q=1$ when the spin current is turned off (see Fig.~\ref{FIG1}(b)).
The stability diagram has been explored in Ref.~\onlinecite{Sampaio} in the absence of the spin current.
For instance, it is stable for $D=4$ mJ m$^{-2}$ and $B_{z}=0$ mT, while it is unstable for $D=2$ mJ m$^{-2}$ and $B_z=50$ mT.
Indeed, when the spin current is switched off, a magnetic skyrmion with $Q=1$ remains stable or is destroyed according to these parameter choices, as depicted in cyan or in blue in Fig.~\ref{FIG1}(b).

\section{Theoretical analysis of the high-Q skyrmion}
\subsection{Hamiltonian}
The Hamiltonian of the system is given by
\begin{align}
H=& -J\sum_{\langle i,j\rangle }\boldsymbol{m}_{i}\cdot \boldsymbol{m}_{j}
+\sum_{\langle i,j\rangle }\boldsymbol{D}\cdot (\boldsymbol{m}_{i}\times \boldsymbol{m}_{j}) \notag \\
& +K\sum_{i}[1-(m_{i}^{z})^{2}]+B_{z}\sum_{i}m_{i}^{z}+H_{\text{DDI}},
\label{HamilAF}
\end{align}
where $\boldsymbol{m}_{i}$ represents the local magnetic moment orientation normalized as $|\boldsymbol{m}_{i}|=1$ at the site $i$, and $\left\langle i,j\right\rangle$ runs over all the nearest neighbor sites in the FM layer. The first term represents the FM exchange interaction with the FM exchange stiffness $J$. The second term represents the DMI with the DMI vector $\boldsymbol{D}$. The third term represents the perpendicular magnetic anisotropy (PMA) with the anisotropic constant $K$. The fourth term represents the Zeeman interaction. The fifth term $\ H_{\text{DDI}}$ represents the dipole-dipole interaction. Although we have included the dipole-dipole interactions in all numerical calculations, the effect is negligible since the size of a magnetic skyrmion is of the order of nanometers.

\subsection{Topological number}
The classical field $\boldsymbol{m}(\boldsymbol{x})$ is introduced for the spin texture in the FM system by considering the zero limit of the lattice constant, that is, $a\rightarrow 0$. The ground-state spin texture is $\boldsymbol{m}=(0,0,1)$. We employ the continuum theory when we make an analytic study of the system.

A magnetic skyrmion is a spin texture which has a topological number. Spins swirl continuously around the core, where spins point downward, and approach the spin-up state asymptotically. The magnetic skyrmion is characterized by the topological number known as the Pontryagin number.
It is given by
$\widehat{Q}=\int d^{2}x q(\boldsymbol{x})$ with the density
\begin{equation}
q(\boldsymbol{x})=-{\frac{1}{4\pi }}\boldsymbol{m}(\boldsymbol{x})\cdot \left(
\partial _{x}\boldsymbol{m}(\boldsymbol{x})\times \partial _{y}\boldsymbol{m}(\boldsymbol{x})\right). \label{PontrNumbe}
\end{equation}
The spin configuration of a magnetic skyrmion is parametrized as
\begin{eqnarray}
m_{x} &=&\cos \phi \left( \varphi \right) \sin \theta \left( r\right) ,\quad
m_{y}=\sin \phi \left( \varphi \right) \sin \theta \left( r\right) ,\quad
\notag \\
m_{z} &=&\cos \theta \left( r\right) , \label{Q}
\end{eqnarray}
where $\varphi$ is the azimuthal angle and $r$ is the radius in the polar coordinate. The topological charge density $q(\boldsymbol{x})$ is shown to be a total derivative, and hence the topological number is a boundary value. It is explicitly calculated as
\begin{equation}
\widehat{Q}=\frac{1}{4\pi}[\cos\theta(\infty)-\cos\theta(0)][\phi(2\pi)-\phi(0)],
\end{equation}
which does not depend on the detailed profile of $\cos\theta (r)$ and $\phi(\varphi)$. The boundary conditions $\cos\theta(0)=-1$ and $\cos\theta(\infty)=1$ are imposed for any skyrmion at the skyrmion center ($r=0$) and at infinity ($r=\infty$). When $\phi=Q\varphi +\chi$, the topological number is $Q$, where $\chi$ stands for the helicity. $Q$ must be an integer for the single-valuedness. In general, $\theta$ and $\chi$ are functions of time $t$. The latter interpolates the N{\'{e}}el-type ($\chi=0,\pi$) or Bloch-type ($\chi=\pi /2,3\pi /2$) skyrmion. We show the spin configurations of the magnetic skyrmions with $Q=1$ and $Q=2$ in Fig.~\ref{FIG1}(a). Spins rotate $Q$ times as $\varphi$ changes from $\varphi=0$ to $\varphi=2\pi$ for the magnetic skyrmion with $Q$.
That is to say, when going around the spin texture of the magnetic skyrmion, the in-plane component of the spin rotates by $2\pi Q$.

\begin{figure*}[t]
\centerline{\includegraphics[width=1.00\textwidth]{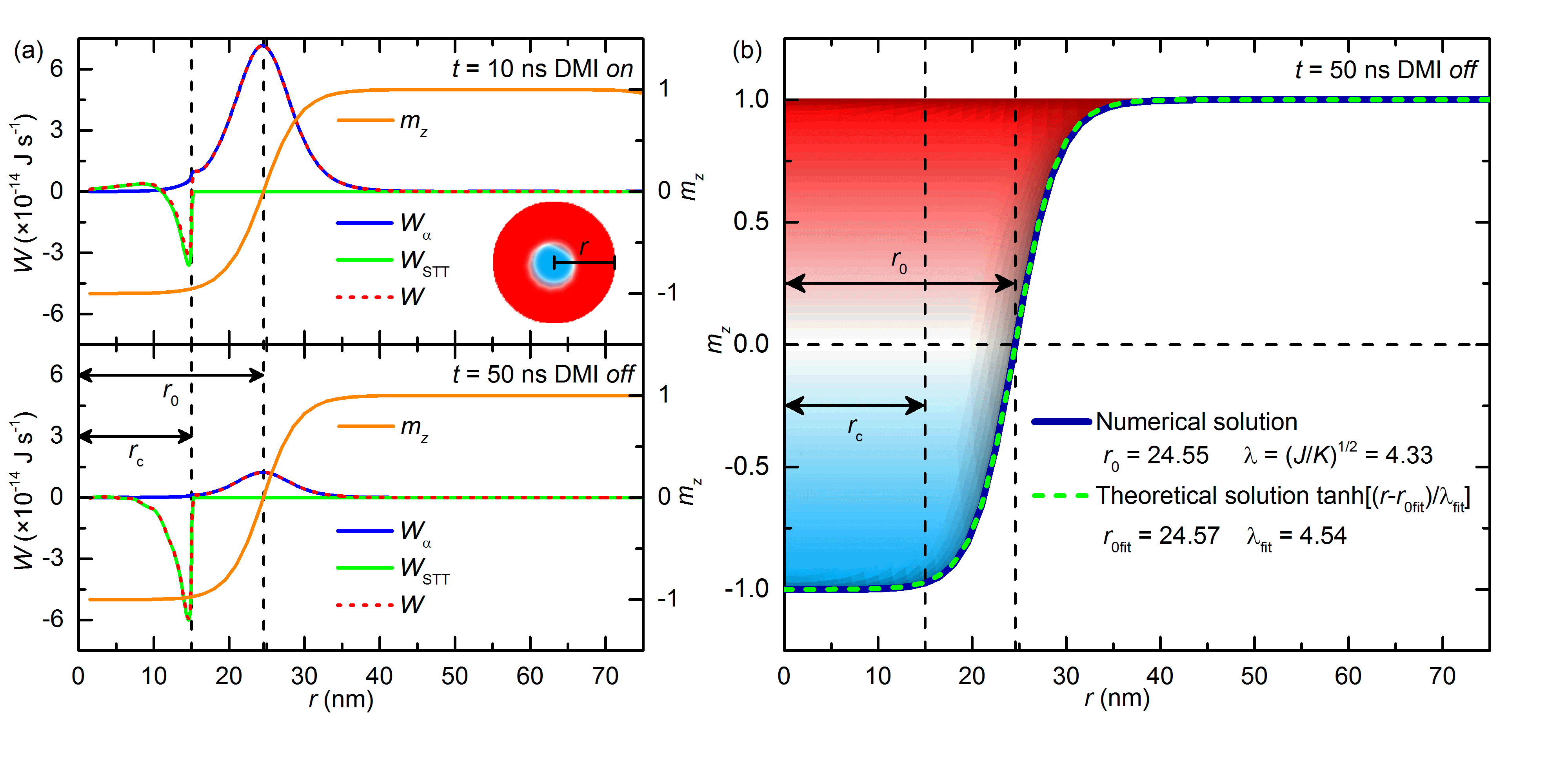}}
\caption{
(Color online)
Profile of $m_{z}$ and the Rayleigh dissipation function $W$ of the high-Q skyrmion with $Q=2$ in the presence and absence of the DMI.
(a) The radius of the FM nanodisk equals $75$ nm, and the spin-polarized current is injected into a circle region with a radius of $r_{c}=15$ nm. The simulated skyrmion radius $r_{0}$ is equal to $24.55$ nm, which is defined as the radius of the circle where $m_{z}=0$.
The DMI constant $D=2$ mJ m$^{-2}$. The spin current density $j=3\times 10^{12}$ A m$^{-2}$. The external magnetic field $B_z=250$ mT.
(b) The form of $m_{z}(r)$ is fitted by the domain wall solution equation~(\ref{tanh}) with the use of $\lambda=4.54$ nm, which is in good agreement with the theoretical value $\lambda=\sqrt{J/K}=4.33$ nm. The functions $W_{\alpha}$ and $W_{\text{STT}}$ in (a) are also well fitted by the same domain-wall solution.
}
\label{FIG7}
\end{figure*}

\begin{figure}[t]
\centerline{\includegraphics[width=0.50\textwidth]{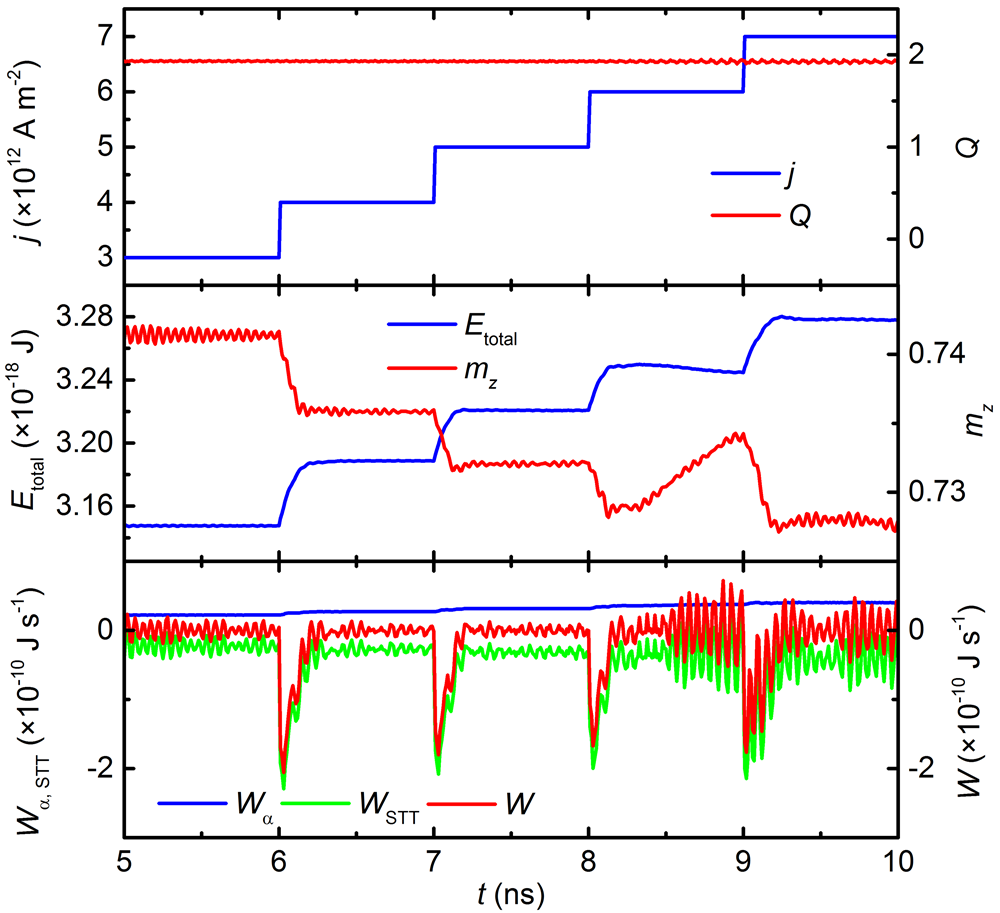}}
\caption{
(Color online)
The high-Q skyrmion with $Q=2$ under stepwise increasing of the spin-polarized current injection.
We show how the topological number $Q$, the total energy $E_{\text{total}}$, the averaged out-of-plane magnetization $m_{z}$, and the dissipation functions ($W_{\alpha}$, $W_{\text{STT}}$, $W$) change, when the injected spin current density $j$ is increased.
The DMI constant $D=2$ mJ m$^{-2}$. The external magnetic field $B_z=250$ mT.
The topological number $Q$ remains as it is when we change the spin current density. The average of $W$ is zero, which implies the energy is balanced in average. It demonstrates that the topological protection against the change of a variable, that is, the spin current injection.
}
\label{FIG8}
\end{figure}

\begin{figure*}[t]
\centerline{\includegraphics[width=1.00\textwidth]{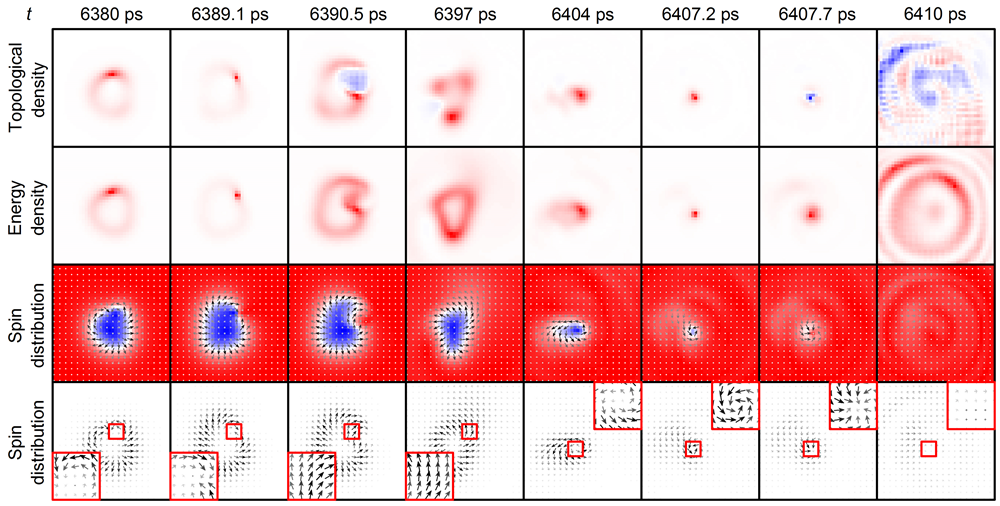}}
\caption{
(Color online)
Annihilation process of the high-Q skyrmion with $Q=2$.
Snapshots of the topological charge density $q(\mathbf{x})$, the energy density $\epsilon(\mathbf{x})$, the spin-component density $m_z(\mathbf{x})$, and its close up at sequential times.
The DMI constant $D=2$ mJ m$^{-2}$. The external magnetic field $B_z=250$ mT. The spin current density is switched off at $t=5000$ ps.
In the simulation, each cell corresponds to one spin, and the cell size is $1.5$ nm $\times$ $1.5$ nm $\times$ $1$ nm.
In the spin distribution panels, each arrow stands for four spins, while it stands for one spin in the insets.
In the first step, the topological number changes from $Q=2$ to $Q=1$, where the skyrmion size remains almost unchanged. Then, the magnetic skyrmion with $Q=1$ shrinks to the size of the lattice scale.
In the second step, the topological number changes from $Q=1$ to $Q=0$, and the magnetic skyrmion disappears.
\blue{See Ref.~\onlinecite{SI} for Supplementary Movie 5.}
}
\label{FIG9}
\end{figure*}

\subsection{Dzyaloshinskii-Moriya interaction}
The DMI is the N{\'{e}}el-type or Bloch-type depending on whether it is introduced from the surface or bulk. We take the interface-induced N{\'{e}}el-type DMI,
\begin{equation}
H_{\text{DM}}=D_{\perp }\int \!d^{2}x\left[ n_{z}\text{div}\mathbf{n}-(\mathbf{n}\cdot \mathbf{\nabla })n_{z}\right] .
\end{equation}
We substitute the magnetic skyrmion configuration equation~(\ref{Q}) into the DMI Hamiltonian, and we find
\begin{eqnarray}
H_{\text{DM}} &=&D_{\perp }\int \!rdrd\varphi \frac{1}{2r}\cos \left[ \left(
Q-1\right) \varphi +\chi \right] \notag \\
&&\times \left( Q\sin 2\theta \left( r\right) +2r\partial _{r}\theta \left(
r\right) \right).
\end{eqnarray}
For $Q\neq 1$, by integrating over $\varphi$, we find $H_{\text{DM}}=0$. As a result, the DMI does not prevent a static magnetic skyrmion from shrinking to a point unless $Q=1$. Hence, there is no static magnetic skyrmion stabilized by the DMI for $Q\neq 1$.

\subsection{Rayleigh dissipation function}
The system contains an energy injection by the spin-polarized current and an energy dissipation by the Gilbert damping. They cannot be analyzed in the framework of the Hamiltonian formalism, where the energy is a constant of motion. It is described by the generalized Lagrangian formalism including the Rayleigh dissipation function.

The Rayleigh dissipation function consists of two terms, $W=W_{\alpha}+W_{\text{STT}}$, the Gilbert damping term~\cite{Tatara},
\begin{equation}
W_{\alpha }=\hbar\alpha \left( \frac{d\mathbf{m}}{dt}\right) ^{2}=\hbar\alpha \left(
\dot{\theta}^{2}+\dot{\phi}^{2}\sin ^{2}\theta \right) , \label{Gilbert}
\end{equation}
with the Gilbert damping constant $\alpha$, and the STT term,
\begin{equation}
W_{\text{STT}}=\hbar |\gamma |u\mathbf{z}\cdot \left( \mathbf{\dot{m}}\times
\mathbf{m}\right) =-\hbar |\gamma |u\dot{\phi}\sin ^{2}\theta , \label{STT}
\end{equation}
where $u$ describes the injection of the spin-polarized current, $u\left(r\right) =|\frac{\hbar}{\mu_{0}e}|\frac{P}{2dM_{S}}j\left(r\right)$ with $j\left(r\right)$ representing the injected current and $\mathbf{z}=\left(0,0,1\right)$. We take $u(r)=u_{0}$ for $r<r_{c}$ and $u(r)=0$ for $r>r_{c}$. We note that $W_{\alpha}>0$, while $W_{\text{STT}}$ can be positive or negative depending on the direction of the spin current. We use the first (second) equations in equation~(\ref{Gilbert}) and equation~(\ref{STT}) for numerical (analytical) calculations.

The generalized Lagrange equation reads~\cite{Goldstein}
\begin{equation}
\frac{d}{dt}\frac{\delta L}{\delta \dot{\mathcal{Q}}}-\frac{\delta L}{\delta \mathcal{Q}}=-\frac{\delta W}{\delta \dot{\mathcal{Q}}},
\end{equation}
where $L$ is the Lagrangian and
$\mathcal{Q}$ is the generalized coordinate.
By taking $\mathbf{m}$ as $\mathcal{Q}$, the generalized Lagrange equation yields the Landau-Lifshitz-Gilbert-Slonczewski (LLGS) equation,
\begin{equation}
\frac{d\mathbf{m}}{dt}=-|\gamma |\mathbf{m}\times \mathbf{B}^{\text{eff}}
+\alpha \mathbf{m}\times \frac{d\mathbf{m}}{dt}+|\gamma |u\mathbf{m}\times
\left( \mathbf{z}\times \mathbf{m}\right) , \label{LLGS}
\end{equation}
with $\hbar\mathbf{B}^{\text{eff}}=-\partial H/\partial \mathbf{m}$.

The energy $E$ changes in the presence of the energy injection and dissipation, $dE/dt=-2\int d^{2}xW\not=0$, in general. Nevertheless, when we take the time average, we should have
\begin{equation}
\int d^{2}x\langle W\rangle =0, \label{weak}
\end{equation}
because this is necessary for a dynamically stabilized magnetic skyrmion. We may call it the weak stationary condition.

\subsection{Skyrmion solution}
We substitute equation~(\ref{Q}) in the LLGS equation~(\ref{LLGS}), which leads to a set of two equations for $\theta (t,r,\varphi)$ and $\chi (t,r,\varphi)$. They are too complicated to solve, reflecting complicated behaviors revealed by numerical solutions (\blue{see Ref.~\onlinecite{SI} for Supplementary Movies 1-3}). However, when we set $D_{\perp}=0$, simple behaviors have been revealed by numerical simulations. Hence, we solve them by setting $D_{\perp}=0$ as the unperturbed system. 

We search for a solution such that $\dot{\theta}=0$. By substituting equation~(\ref{Q}) in the LLGS equation~(\ref{LLGS}), and by setting $\dot{\theta}=0$, we obtain
\begin{align}
-J(\partial _{r}^{2}\theta & +\frac{\partial _{r}\theta }{r})+(\frac{JQ^{2}}{2r^{2}}+K)\sin 2\theta  \notag \\
& +B_{z}\cos \theta +\frac{1}{\alpha }u(r)\sin \theta =2D_{\perp }F(r,\theta
,\varphi ),  \label{BasicEq}
\end{align}
where
\begin{align}
F=&-\frac{Q}{r}\sin \theta \cos [(1-Q)\varphi +\chi ] \notag\\
&-\frac{1}{\alpha }\partial_{r}\theta \sin [(1-Q)\varphi +\chi ]\sin \theta . \label{EqF}
\end{align}
The role of the injected spin-polarized current ($u\not=0$) is to impose the boundary condition $\cos\theta=-1$ at $r=0$.

When $Q=1$, since $F=-\sin \theta /r$, equation~(\ref{BasicEq}) is numerically solvable with respect to $\theta$ with the boundary condition $\cos\theta=-1$ at $r=0$ and $\cos\theta=1$ at $r=\infty$. The azimuthal angle is given by $\varphi _{0}(t)=(\gamma u/\alpha Q)t$. The equations of motion are well approximated by $J\partial_{r}^{2}\theta=K\sin 2\theta$ for any $Q$ asymptotically. This equation has the domain-wall solution, 
\begin{equation}
\cos \theta =\tanh \frac{r-r_{0}}{\lambda }, \label{tanh}
\end{equation}
with the domain-wall width $\lambda=\sqrt{J/K}$ and the skyrmion radius $r_{0}$.

Our major interest is the case of $Q\neq 1$. Since $\theta$ and $\varphi$ are coupled, it is not straightforward to solve for a magnetic skyrmion. Let us require the weak stationary condition by taking the time average of equation~(\ref{EqF}). We find that $\langle F\rangle=0$ unless $Q=1$. Then, equation~(\ref{BasicEq}) is solvable with respect to $\theta$ with the boundary condition $\cos\theta=-1$ at $r=0$ and $\cos\theta=1$ at $r=\infty$.

The profile of $m_{z}(r)$ is given by equation~(\ref{Q}) and equation~(\ref{tanh}). The Gilbert damping term $W_{\alpha}(r)$ and the STT term $W_{\text{STT}}(r)$ are given by equations~(\ref{Gilbert})~-~(\ref{STT}) together with $\dot{\theta}=0$, $\dot{\phi}=\text{constant}$ and equation~(\ref{tanh}) or $\sin^{2}\theta=\sec^{2}[(r-r_{0})/\lambda]$. The theory and numerical simulation lead to identical results which overlap within the precision of numerical simulation as shown in Fig.~\ref{FIG7}(b).

On the other hand, when the injected spin-polarized current is switched off, that is, $u=0$, there is no skyrmion solution and we recover the FM ground state $\cos\theta=1$.

\section{Conclusions}
We have analyzed the nucleation, the stability, and the destruction of the high-Q skyrmion with $Q=2$ in ferromagnets.
It is realized when the in-plane component of the spin rotates by $2\pi Q$ and the magnetic skyrmion has acquired a high helicity $Q$.
In the presence of the DMI, although there exist static magnetic skyrmions with $Q=1$, there exist no static high-Q skyrmions.
Nevertheless, a high-Q skyrmion can be created and stabilized dynamically by injecting the spin-polarized current. The DMI plays a crucial role in the creation mechanism by twisting spins and generating fluctuations of the energy density and the topological charge density.
Indeed, we have numerically verified that a high-Q skyrmion cannot be created dynamically without the DMI.

We have also observed numerically that a high-Q skyrmion can be created in a wide range of parameters as well as the spin current injection size.
Furthermore, the high-Q skyrmion has been found to be created and maintained even at finite temperature, although its structure is deformed due to thermal fluctuations.

The nucleation process of a high-Q skyrmion is revealed by investigating the magnetization distribution, the topological charge density, and the energy density. It occurs in two steps.
First, it so happens that the high density part is localized to a lattice-scale area in the boundary of a magnetic bubble with $Q=0$. The topological number jumps from $Q=0$ to $Q=1$ by making a large spin rotation in this lattice-scale area within a few picoseconds. This phenomenon would be viewed as an emergence of a Bloch point in the continuum theory.
Second, a similar process occurs to make a jump from $Q=1$ to $Q=2$. The dissipation spreads over the sample like a burst at the moment of the birth of a magnetic skyrmion.

The continuity of the spin texture is recovered since a smooth texture has a lower energy. Once a sufficiently smooth high-Q skyrmion is generated, it remains stable even if we switch off the DMI or even if we change the current density of the injected spin current considerably. We have explained its stability as a topologically protected dissipative structure.

When the spin current injection is switched off, the high-Q skyrmion is destroyed. The destruction process occurs also in two steps as in the case of the nucleation process. However, the detailed mechanism is different.
The first step occurs by way of a Bloch point, where the continuum picture is still good.
In the second step, the lattice structure becomes important since the skyrmion size is so small, where the skyrmion spin texture turns into the vortex spin texture and disappears. The dissipation spreads over the sample like a burst at the moment of the destruction of a magnetic skyrmion.

It is a hard problem to solve the nucleation or destruction process analytically since it is a highly nonlinear process involving Bloch points. Furthermore, the lattice structure plays a key role in these processes at the microscopic level.
We hope this work provides useful guidelines in searching new type of skyrmions and will afford a new dimension towards fully understanding the nontrivial topology of magnetic skyrmions with higher topological numbers. 

\begin{acknowledgments}
Y.Z. acknowledges the support by National Natural Science Foundation of China (Project~No.~1157040329), the Seed Funding Program for Basic Research and Seed Funding Program for Applied Research from the HKU, ITF Tier 3 funding (ITS/171/13 and ITS/203/14), the RGC-GRF under Grant HKU 17210014, and University Grants Committee of Hong Kong (Contract~No.~AoE/P-04/08).
M.E. thanks the support by the Grants-in-Aid for Scientific Research from MEXT KAKENHI (Grant~Nos.~25400317~and~15H05854).
%
%
M.E. is very much grateful to N. Nagaosa for many helpful discussions on the subject.
X.Z. was supported by JSPS RONPAKU (Dissertation Ph.D.) Program.
X.Z. greatly appreciates on-going discussions with J. Xia.
\end{acknowledgments}

\appendix*
\section{SIMULATION AND MODELING}
The micromagnetic simulation is carried out with the well-established Object Oriented MicroMagnetic Framework (OOMMF) software (1.2a5 release)~\cite{OOMMF}.
The OOMMF extensible solver (OXS) extension module of the interface-induced DMI, that is, the Oxs{\_}DMExchange6Ngbr class, is included in the simulation~\cite{Rohart}.
The OXS extension module of the thermal fluctuation, that is, the Xf{\_}ThermSpinXferEvolve class, is employed to simulate the finite-temperature system.
The 3D time-dependent magnetization dynamics at zero temperature is determined by the LLGS equation~\cite{OOMMF,LLGSTT}, while a highly irregular fluctuating field representing the irregular influence of temperature is added into the LLGS equation for simulating the magnetization dynamics at finite temperature. The average energy density of the system contains the exchange energy, the anisotropy energy, the applied field (Zeeman) energy, the DMI energy, and the magnetostatic (demagnetization) energy terms.

In the simulation, we consider a $1$-nm-thick FM nanodisk with a diameter of $150$ nm, which is attached to a heavy-metal substrate.
The material parameters used by the simulation program are adopted from Refs.~\onlinecite{Fert,Sampaio}:
the Gilbert damping coefficient $\alpha=0.01$,
the gyromagnetic ratio $\gamma=-2.211\times 10^{5}$ m A$^{-1}$ s$^{-1}$,
the saturation magnetization $M_{S}=580$ kA m$^{-1}$,
the exchange stiffness $J=15$ pJ m$^{-1}$,
the interface-induced DMI constant $D=0\sim 3$ mJ m$^{-2}$,
and the PMA constant $K=0.8$ MJ m$^{-3}$.
The polarization rate of the vertical spin current applied in the simulation is fixed at $P=0.4$.
The simulated model is discretized into tetragonal cells with the optimum cell size of $1.5$ nm $\times$ $1.5$ nm $\times$ $1$ nm, which gives a good trade-off between the computational accuracy and efficiency.
The finite-temperature simulation is performed with a fixed time step of $1\times10^{-14}$ s, while the time step in the simulation with zero temperature is adaptive ($\sim 6\times 10^{-14}$ s). The seed value in the finite temperature simulation is fixed at $100$. The output is made with an interval of $1\times 10^{-13}\sim 1\times 10^{-11}$ s.
The initial magnetization distribution of the FM nanodisk is relaxed along the $+z$-direction, except for the tilted magnetization on the edge due to the DMI. The external magnetic field is applied perpendicular to the FM nanodisk pointing along the $+z$-direction with an amplitude of $B_{z}=0\sim 500$ mT.

\end{document}